\documentclass[useAMS,usenatbib,11pt]{mn2e}

\usepackage{graphicx}
\usepackage{verbatim}
\usepackage{bm}
\usepackage{mathptmx}

\DeclareMathSymbol{\bolditalicgamma}{\mathalpha}{letters}{"0D}

\title[Improving the precision of pulsar timing]{Improving the precision of pulsar timing through polarization statistics}

\author[S. Os{\l}owski et al.]{S. Os{\l}owski$^{1,2}$\thanks{E-mail:soslowski@astro.swin.edu.au}, W. van Straten$^{1}$,   P. Demorest$^{3}$ and M. Bailes$^{1}$\\
$^{1}$Swinburne University of Technology, Centre for Astrophysics and Supercomputing, Mail H39, PO Box 218, VIC 3122, Australia\\
$^{2}$CSIRO Astronomy and Space Sciences, Australia Telescope National Facility, P.O. Box 76, Epping, NSW 1710, Australia\\
$^{3}$National Radio Astronomy Observatory, 520 Edgemont Road, Charlottesville, Virginia 22093, USA\\}

\begin{document}

\newcommand{\psr}{\mbox{PSR J0437$-$4715}}
\newcommand{\wvect[1]}{{\bsf #1}}
\newcommand{\bvect[1]}{{\bsf #1}$_{\bm 0}$}

\let\oldhat\hat
\renewcommand{\vec}[1]{\mathbf{#1}}
\renewcommand{\hat}[1]{\oldhat{\mathbf{#1}}}
\renewcommand{\S}[0]{Section }

\date{Accepted . Received ; in original form}

\pagerange{\pageref{firstpage}--\pageref{lastpage}} 
\pubyear{2012}

\maketitle

\label{firstpage}
  
\begin{abstract}

At the highest levels of pulsar timing precision achieved to date, experiments are limited by noise intrinsic to the pulsar. This stochastic wideband impulse modulated self-noise limits pulsar timing precision by randomly biasing the measured times of arrival and thus increasing the root mean square (rms) timing residual. We discuss an improved methodology of removing this bias in the measured times of arrival by including information about polarized radiation. Observations of \psr\ made over a one-week interval at the Parkes Observatory are used to demonstrate a nearly 40 per cent improvement in the rms timing residual with this extended analysis. In this way, based on the observations over a 64 MHz bandwidth centred at 1341 MHz with integrations over $16.78$~s we achieve a 476 ns rms timing residual. In the absence of systematic error, these results lead to a predicted rms timing residual of 30 ns in one hour integrations; however the data are currently limited by variable Faraday rotation in the Earth's ionosphere. The improvement demonstrated in this work provides an opportunity to increase the sensitivity in various pulsar timing experiments, for example pulsar timing arrays that pursue the detection of the stochastic background of gravitational waves. The fractional improvement is highly dependent on the properties of the pulse profile and the stochastic wideband impulse modulated self-noise of the pulsar in question.

\end{abstract}


\section{Introduction}
\label{Section::intro}

Radio pulsars are very stable rotators whose radio emission is received on Earth as series of radio pulses. The measurement of the average pulse train time of arrival (ToA) is referred to as pulsar timing and leads to the majority of derived pulsar astrophysics. At the heart of this methodology lies the comparison of the estimated ToA to a model of the pulsar's properties, including astrometric, spin-down, and binary (both Keplerian and relativistic) parameters. Millisecond pulsars (MSPs), with their short periods, narrow features in their phase-resolved light curves (pulse profiles) and very low spin frequency derivatives, are exceptionally stable \citep{1997A&A...326..924M} and provide the best opportunity for high precision timing experiments. A number of authors have achieved timing precision of a few hundred nanoseconds over time-scales of five to ten years \citep[e.g.,][and references therein]{2009MNRAS.400..951V}. With such high precision, a number of experiments are possible, such as testing theories of relativistic gravity \citep[e.g.][]{1975ApJ...195L..51H,2005ASPC..328...25W,2006Sci...314...97K}; detection of planets orbiting a pulsar \citep{1992Natur.355..145W,2011Sci...333.1717B} and the most precise published mass measurements of planets in the Solar System \citep{2010ApJ...720L.201C}; determining properties of the interstellar medium \citep[ISM, e.g.][]{2007MNRAS.378..493Y,2011MNRAS.416.2821D}; constraining the equation of state of dense matter \citep[e.g.][]{2010Natur.467.1081D}; or detecting irregularities in terrestrial time-scales and developing a pulsar based one \citep[][]{1996A&A...308..290P,2008MNRAS.387.1583R,2012MNRAS.427.2780H}.

One particular application of the pulsar timing methodology is the detection of gravitational waves, i.e., perturbations of the space-time metric predicted by general relativity. These waves carry energy and so their emission can be inferred by observing the decrease of the semi-major axis of an emitting binary system of the order of $1$~cm a day, as detected with pulsar timing of binaries \citep[e.g.,][]{1982ApJ...253..908T,1998ApJ...505..352S,2006Sci...314...97K,2006AAS...209.9101J,2008PhRvD..77l4017B,2012MNRAS.423.3328F}. A method of  direct detection of gravitational waves relies on their impact on the relative distance between Earth and the pulsar with a quadrupolar spatial signature as the gravitational wave passes through space-time \citep[e.g.,][]{2006ApJ...653.1571J}.

A number of collaborations around the world are observing $\sim30$ millisecond pulsars in the pursuit of detecting gravitational waves with so called pulsar timing array experiments. These experiments, based on early ideas of \citet{1978SvA....22...36S} and \citet{1990ApJ...361..300F}, exploit the correlation between the timing residuals for pairs of pulsars in different directions \citep{1983ApJ...265L..39H} in the sky to measure the spatial signature of the gravitational wave background and thus distinguish it from the timing noise, interstellar medium (ISM) effects, clock or Solar system ephemeris errors, which have different multipole signatures or are uncorrelated between different objects. We note that recently \citet{vanStraten2012} pointed out that instrumental distortions due to imperfect polarimetric calibration can corrupt the quadrupolar signature of the gravitational waves.

A number of authors have demonstrated that in order to detect gravitational waves, a high precision of pulsar timing observations must be achieved. For example, \citet{2006ApJ...653.1571J} showed that 20 pulsars need to be timed at a precision of $100$ to $500$~nanoseconds over a period of 10 years. While the desired precision has been attained for a small number of objects, the timing precision of the MSPs is worse than the theoretically predicted values for objects expected to be the best timers. A prime example of this is the closest and brightest MSP, \psr\ \citep{1993Natur.361..613J}. Despite numerous attempts, the timing precision of this pulsar has been significantly worse than expected from the pulse profile and signal-to-noise ratio. \citet{1997ApJ...478L..95S}, \citet{2001Natur.412..158V} and \citet{2008ApJ...679..675V} all achieved progressively better timing precision with the latest results being $199$\,ns rms timing residual over 10 years with integration lengths between $10$ and $120$ minutes, still a factor of 4 or 5 worse than theoretical predictions. An important factor in the improvement of the timing precision was related to improved methods of dealing with  polarimetric calibration \citep{2000ApJ...532.1240B,2004ApJS..152..129V,2006ApJ...642.1004V}.

The long standing problem of \psr\ under-performing in terms of timing precision is now understood \citep[][paper~I hereafter]{2011MNRAS.418.1258O} which demonstrated that the theoretical predictions neglect the contribution of the stochastic wideband impulse modulated self-noise (SWIMS), also referred to as pulse (or phase) jitter \citep[e.g.,][]{2010arXiv1010.3785C,2012MNRAS.420..361L}. For pulsars whose peak flux approaches the system equivalent flux density this contribution cannot be neglected and it biases the estimates of the times of arrival (ToA) of the pulse train. Paper~I also presented a method based on decomposition of the residual profiles onto eigenvectors of the noise covariance matrix to help remove the effects of SWIMS in post processing of timing residuals. We applied this method to \psr\ and achieved a 20 per cent reduction in the rms timing residual. The current work is concerned with improvements to this methodology, allowing even better recovery of unbiased ToAs by analysing the polarized flux of the pulsar.

This paper is structured as follows: \S \ref{Section::selfnoise} introduces the concept of stochastic wideband impulse modulated self-noise and its impact on attainable timing precision. \S \ref{Section::observations} describes the data used and details of processing techniques applied. In \S \ref{Section::method} we present an extension to the methodology allowing for removal of the bias introduced by SWIMS. The results of applying this method to archival observations of \psr\ are presented in \S \ref{Section::results}. We discuss these results in detail in \S \ref{Section::discussion} before drawing our conclusions in \S \ref{Section::conclusions}.  

The work presented in this paper is a logical extension of the work presented in paper~I, which should be consulted in order to fully understand the current work. In the following section we briefly summarise the key concepts of paper~I for the reader's convenience. 

\section {Pulsar Timing and the Stochastic Wideband Impulse Modulated Self-Noise}
\label{Section::selfnoise}

The pulsar timing methodology is based on cross-correlation of the observed pulse profile with a template, typically done in the Fourier domain \citep{1992RSPTA.341..117T}. The derived ToAs  and the pulsar model are used to form timing residuals as implemented by the \textsc{tempo2} software package \citep{2006MNRAS.369..655H}. As outlined in \S \ref{Section::intro}, in the case of \psr\ the rms timing residual is much larger than what is theoretically expected based on the considerations of the pulse profile effective width and the signal-to-noise ratio (S/N) of the data. In paper~I we provided evidence that this is due to the presence of SWIMS, i.e., the pulsar itself becomes a source of significant noise that is normally not accounted for in the noise balance and predictions of timing precision.

The amplitude of SWIMS is phase dependent and thus the variance of noise is also phase dependent, that is the noise is heteroscedastic. In addition, due to the emission properties of the pulsar, SWIMS exhibits temporal and spectral correlations. Both types of correlation have adverse consequences. The temporal correlation implies that the noise in the pulse profile is not only heteroscedastic but also not independent between different phase bins. While this correlation can amplify the bias introduced by SWIMS into the ToA estimation, it allows for removal of this very bias in post-processing. Due to the wideband nature of the pulsar intrinsic noise the signal at different frequencies is not independent. Therefore the ToA estimated from separate parts of the observing radio band are correlated thus neutralising the benefit of increasing the instrumentation bandwidth. Using wideband receivers is still beneficial for other reasons as it allows to address many issues related to the interstellar medium such as dispersion measure variations or scattering \citep[e.g.,][]{2007MNRAS.378..493Y}. 

The temporal correlations and heteroscedasticity of SWIMS can be fully characterised by the covariance matrix of the pulse profiles after subtracting the template profile that was used during the ToA estimation. This allows us to study the noise properties at any integration time longer than the pulse period as the covariance matrix scales inversely with the integration length and the relative contribution of SWIMS to the covariance matrix is integration length independent.

For a more detailed discussion of SWIMS we refer the reader to \S 2 of paper~I.

\section {Observations and data processing}
\label{Section::observations}

\begin{figure*}
\includegraphics[angle=-90,width=\textwidth]{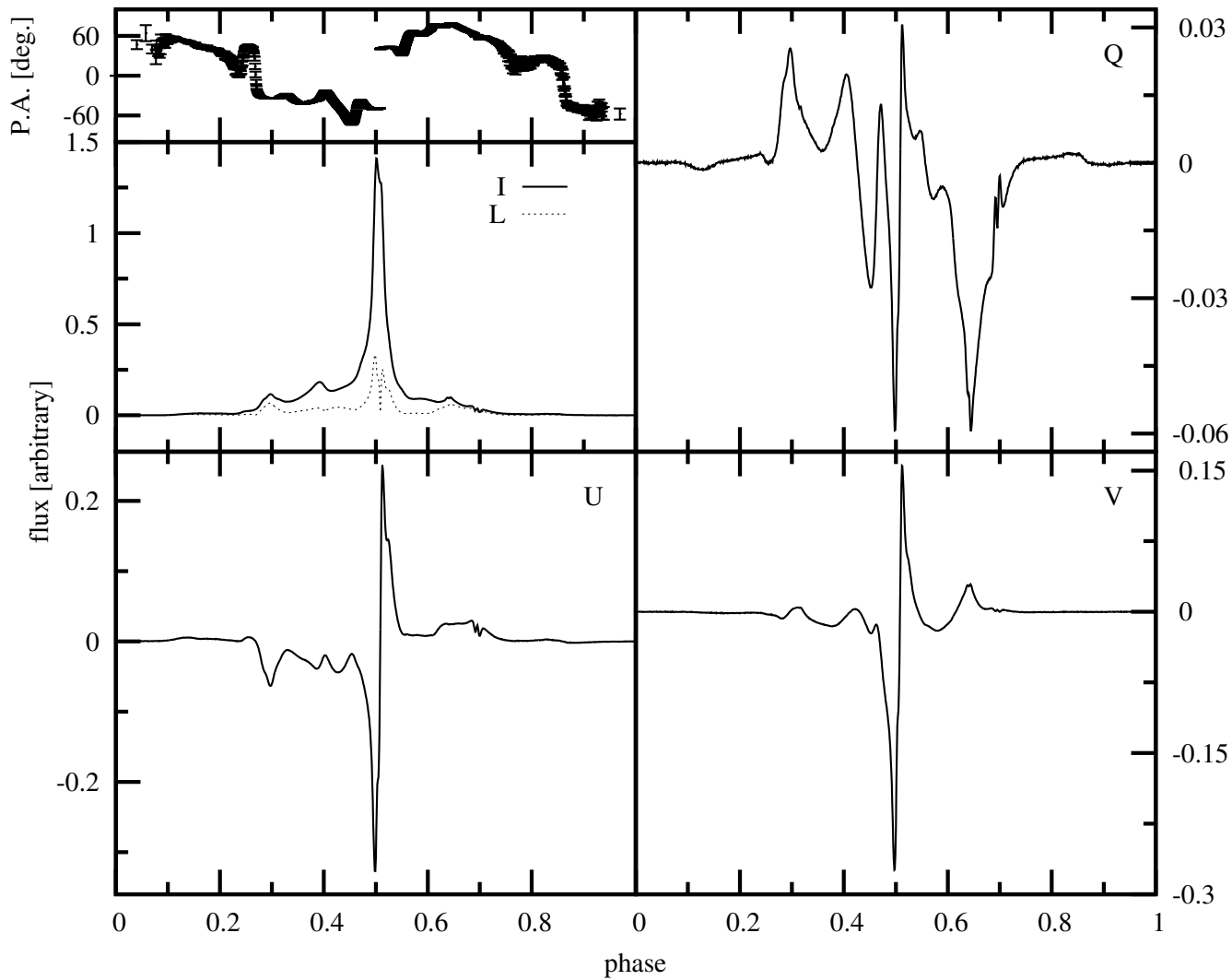}
\caption{A high S/N template of \psr\ in the 20cm observing band formed from 41.7 hours of data. The top left short panel shows the position angle swing and is situated directly above a panel showing the total intensity pulse profile in solid line and the linear polarization in a dotted line. The top right panel shows the Stokes Q parameter of the template, while the bottom left and right panels correspond to Stokes U and V, respectively. The panels on the left use the ordinate axes on the left edge while the right hand side panels use the far right ordinate axes. All stokes parameters are on the same flux scale.
\label{Fig::template_SWIMS_pol}}
\end{figure*}

Dual-polarization data were recorded at the Parkes 64\,m radio telescope using the central beam of the 20\,cm multibeam receiver \citep{1996PASA...13..243S} and the second generation of Caltech Parkes Swinburne baseband recorder \citep[][CPSR2]{2003ASPC..302...57B,2005PhD.....H}. A total of $\sim36.5$~hours were recorded between the 19th and 27th of July 2003 using two bit digitisation with a bandwidth of $64$~MHz centred at 1341 MHz and split over 128 channels, each $0.5$~MHz wide, resolved into 1024 phase bins, and integrated over $16.78$~seconds. CPSR2 adjusts the sampling thresholds twice every minute to maintain linearity of the digitiser response. The incident radiation was dedispersed coherently to preserve the narrow features of the pulse profile. The data were processed as in \citet{vanStraten2012} and we refer the reader to this work for details. The Stokes parameters in the calibrated data follow the conventions defined by \citet{2010PASA...27..104V}. Here, we present a brief summary of the applied processing.

As explained in paper~I and \S \ref{Section::discussion_method} below, it is beneficial to have a large number of sub-integrations available for the analysis. Therefore, we demonstrate our methodology using a different dataset than in paper~I. We describe here an archival dataset ample enough for our purposes. This dataset can be obtained from the CSIRO's Data Access Portal\footnote{http://datanet.csiro.au/dap/} \citep{2011PASA...28..202H}.

We generated a high S/N, calibrated full stokes template profile using an independent dataset, as follows: 48 observing sessions of \psr, each at least four hours in duration, spread over 7.2 years were calibrated using the measurement equation modelling technique \citep{2004ApJS..152..129V}. This technique makes use of observations of a pulsed noise diode that injects a polarized reference signal into the feed horn. Observations of the radio source 3C 218 (assumed to have constant flux and negligible circular polarization) were used to constrain the boost along the Stokes V axis. Then, the five best quality of the 8 hour sessions were integrated together to form an initial pulse profile template. The remaining 43 sessions were matched to it using the matrix template matching technique \citep{2006ApJ...642.1004V}. A final pulse profile template was created from the best 99.4 hours of data processed in the same manner and integrated hierarchically to minimise quantisation errors into a profile with S/N equal to $18.6\times10^3$. The full polarization information was retained and is shown in Fig. \ref{Fig::template_SWIMS_pol}. Throughout this publication we use the same template in all contexts.

The generated template is used for polarimetric calibration of the primary dataset spanning a week of observations. Calibration was performed using the measurement equation template matching technique \citep{vanStraten2012}. This method yields superior results by using \psr\ as a polarized reference source. It combines measurement equation modelling with matrix template matching to break the degeneracies described in the Appendix of \citet{2004ApJS..152..129V}.

\begin{figure*}
\includegraphics[angle=0,width=\textwidth]{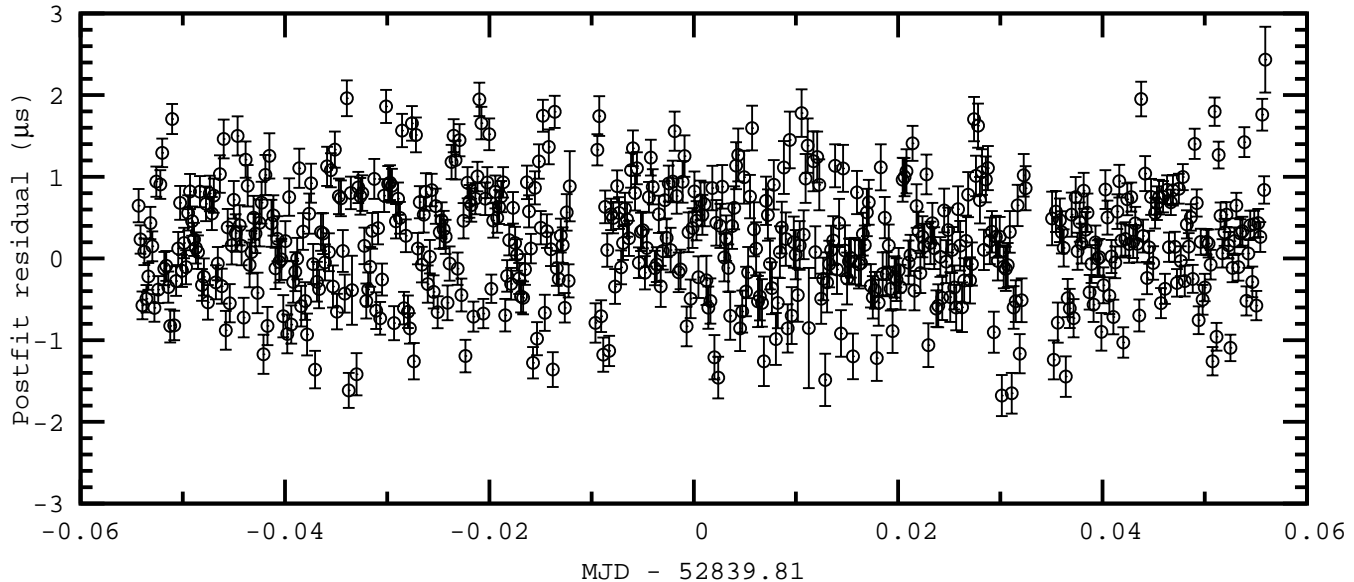}
\caption{Timing residuals estimated for 16.78~s sub-integrations of data taken during one of the days timed against the standard from Fig. \ref{Fig::template_SWIMS_pol}. The mean ToA estimation error in the whole dataset is 255 ns, whereas the weighted rms of the residual ToAs is 774 ns. The fit has $\chi^2/{\rm d.o.f.}$ of 11.8. For clarity, we have plotted the residuals as a function of MJD.
\label{Fig::all_residuals}}
\end{figure*}

In both the dataset used for creating the template profile and the one used for estimating the covariance matrix, 15 per cent of the frequency channels on each side of the band were rejected to avoid quantisation noise and aliasing\footnote{In this context by aliasing we mean reflection of the radio frequencies outside of the band.}. CPSR2 is a 2-bit recorder which therefore introduces significant distortions into the measured pulse profile \citep{1998PASP..110.1467J}. A 2-bit correction scheme \citep{vanStraten2012} has been applied and the data were searched for cases of imperfect correction. We achieve this by measuring the $\chi^2/d.o.f.$ of the fit of the pulse profile to a high signal to noise ratio (S/N) template, where $d.o.f.$ refers to the number of degrees of freedom of the fit. The profiles with worst and uncorrectable 2-bit distortions have distinctly higher values of $\chi^2$ which allows for their easy identification and rejection. All of the processing steps were performed using the {\sc PSRchive}\footnote{http://psrchive.sourceforge.net/} data processing and analysis suite \citep{2004PASA...21..302H,2012arXiv1205.6276V}.

The timing residuals for a subset of the 7845 profiles based on the $16.78$ s sub-integrations used in the analysis are presented in Fig. \ref{Fig::all_residuals} as a function of MJD. The ToAs were estimated using the Fourier domain with Markov chain Monte Carlo algorithm in {\sc PSRchive}. The weighted rms timing residual is $774$~ns and the $\chi^2/{\rm d.o.f.}$  equals $11.8$. The unweighted rms timing residual is slightly higher and equals 806~ns. The mean ToA estimation error is 255 ns, and the mean S/N is 302. Based on the findings of paper~I, we attribute the higher-than-expected rms timing residual to SWIMS and the induced ToA estimation bias and $\chi^2/{\rm d.o.f.}$ on the invalid assumptions used in the ToA uncertainty estimation. The rms timing residual scales as the square root of the inverse of the integration time for pulsars without temporal correlations between subpulses and either without long-term timing noise or when studied over short periods of time \citep[see, e.g.,][]{1975ApJ...198..661H,2011MNRAS.417.2916L,2012ApJ...761...64S}. Based on this extrapolation, which we have shown to be true for \psr\ (see Fig. 3 in Paper I) the data would yield an rms timing residual of 53~ns when derived from 60 minute integrations.

\section{Method}
\label{Section::method}

The method described here is a direct extension of what was presented in \S 4 of paper~I and continues to develop the ideas presented by \citet{2007PhDT........14D}. To summarise, this method consists of fitting for a scale, phase shift and offset between the observed total-intensity pulse profiles and a template profile and calculating the covariance matrix of the residual profiles. Next, the eigenvectors of the covariance matrix are calculated and used as a new basis in which the residual profiles are presented. The eigenvectors corresponding to the highest eigenvalues describe the dominant pulse shape variations. As a final step, a correlation between the new presentation of the residual profiles and timing residuals is found using multiple regression and enables a post-processing correction of the ToA bias.

The method presented in paper~I yielded a 20 per cent reduction of the rms timing residual by exploiting this correlation. However, it is unable to correct the bias introduced by the component of SWIMS that is correlated with any of the three degrees of freedom that are removed from the data during the template-matching fit. Unfortunately, the component of SWIMS correlated with one of these degrees of freedom, namely the template time derivative, i.e., the phase shift, has a high impact on the bias of the ToA. In this work, we extended the previously published method by incorporating the polarized flux of the pulsar. As we demonstrate, this allows to recover some of the signal lost in the previous attempt at the mitigation of additional ToA scatter introduced by SWIMS. 

\begin{figure*}
\includegraphics[angle=-90,width=\textwidth]{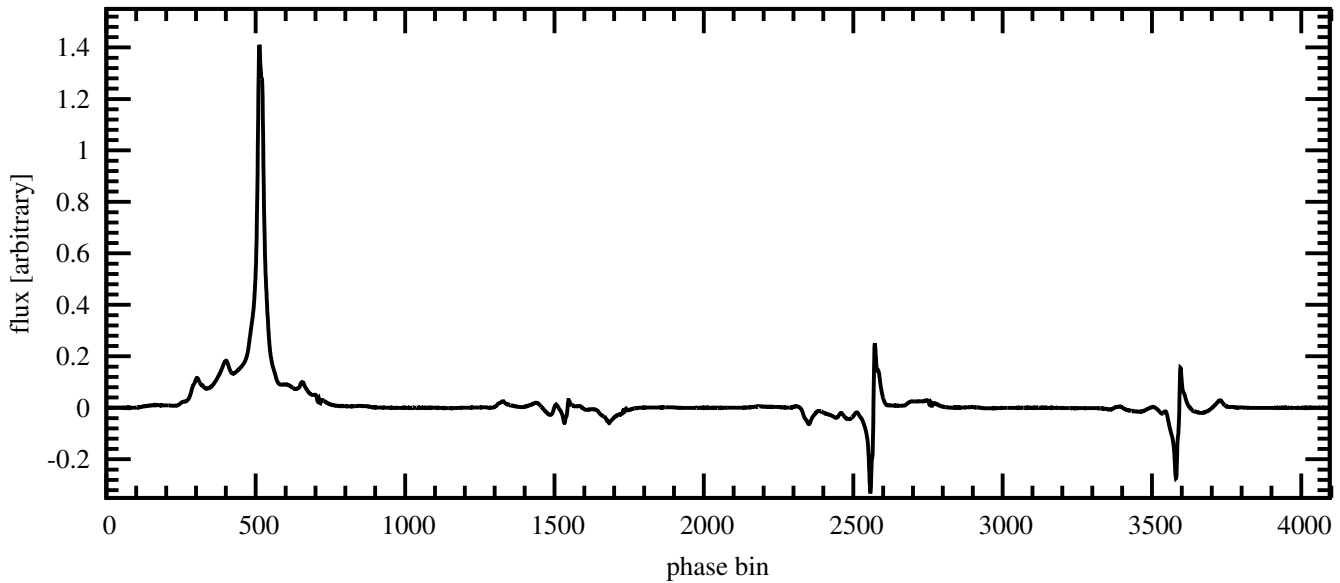}
\caption{An example of a profile used as input for the methodology presented in this work. The profile is formed from all four stokes parameters spread over $4N_{\rm bin}$ phase bins instead of four seperate profiles with $N_{\rm bin}$ phase bins each. We used the same profile as in Fig. \ref{Fig::template_SWIMS_pol}. The input profile is resolved into 1024 pulse phase bins and after combing all the stokes parameters into one profile we obtain a pulse profile with 4096 phase bins.
\label{Fig::fournbin}}
\end{figure*}

The simplest way to implement this extended methodology is by considering the pulse profiles of all four stokes parameters, each resolved in $N_{\rm bin}$ phase bins, as a $4N_{\rm bin}$ profile with only one polarization. The additional, artificially introduced, phase bins are populated with the polarized flux as measured in the stokes parameters. That is, the first $N_{\rm bin}$ phase bins represent the information contained in the total intensity measurement, the bins from $N_{\rm bin}+1$ to $2N_{\rm bin}$ represent the Stokes Q measurements, bins from $2N_{\rm bin}+1$ to $3N_{bin}$ represent the Stokes U flux and the last $N_{\rm bin}$ phase bins are populated with the circular polarization, i.e., the Stokes V parameter. An example of such a profile is shown in Fig. \ref{Fig::fournbin}, where the mean off-pulse values were set equal to zero for each stokes parameter independently.

Before forming the $4N_{\rm bin}$ profiles, we performed template fitting and subtracted the best-fit template from each observation, which is necessary to estimate the residual profile covariance matrix \textbfss{C}. As we considered all four stokes parameters, we have six free parameters of the fit: a scale factor to account for flux variation of the pulsar (intrinsic, related to the effects of interstellar medium, or instrumental effects); a phase shift between the pulse profile and the template; and four offsets between the pulse profile and the template for each of the stokes parameters. The phase shift and scale factor are determined based on the total intensity but applied to all four stokes parameters. A unique baseline offset is calculated from and applied to each of the stokes parameters. The fitting procedure results in six out of $4N_{\rm bin}$ eigenvectors sharing the eigenvalue equal zero and form what we refer to as the fit-space.

After computing the post-template-fit residual we form the $4N_{\rm bin}$ profiles and directly follow the method presented in paper~I in equations 2 through to 11\footnote{Equation 9 in paper~I is missing a multiplicative factor of $1/N$.}. When exploiting the additional information available in all the stokes parameters, the sizes of some of the mathematical quantities involved are different than presented in Paper I, namely: the covariance matrices \textbfss{C} and \textbfss{D} are both $4N_{\rm bin}$ by $4N_{\rm bin}$ square matrices; there are $4N_{\rm bin}$ eigenvalues associated with as many eigenvectors, each of length $4N_{\rm bin}$;  the matrix of projections of residual profiles onto eigenvectors has now $4N_{\rm bin}$ columns; the vector of covariances between the arrival time residuals and the projection coefficients $\mathit{\bgamma}$ has $4N_{\rm bin}$ elements; and there are $4N_{\rm bin}$ values of $\xi$, the Pearson's product moment correlation coefficients between the timing residuals and the projection coefficients onto one of the eigenvectors. After the analysis is finished the $4N_{\rm bin}$ profiles and eigenvectors are converted back to four $N_{\rm bin}$ profiles, one for each of the four stokes parameters. 

Similar to the methodology described in paper~I, we have performed a number of simulations to ensure that the extended methodology does not remove arbitrary phase shifts not related to pulse profile deviation from the template profile. We refer the readers to our previous work for a description of how such simulations are realised. Here we report that no removal of arbitrary shift occurs and thus the method is valid and useful for astrophysical experiments.

An implementation of this method is publicly available as part of {\sc PSRchive} as an upgraded version of the previously available application ``{\sc psrpca}''. It requires the GNU Scientific Library\footnote{http://www.gnu.org/software/gsl/} to work and can be significantly sped up by using the {\sc cula} library\footnote{http://www.culatools.com/} \citep{CULA} if an NVIDIA\textsuperscript{\textregistered} CUDA\textsuperscript{\texttrademark} enabled Graphics Processing Unit is available.

\section{Results}
\label{Section::results}
Applying our method to the observed dataset of 7845 profiles leads to the following:
\begin{itemize}
\item the detection of significant pulse shape variations with at least 21 significant eigenvectors,
\item a reduction in rms timing residual from $776$~ns to $473$~ns, a $39$ per cent reduction; and a
reduction in $\chi^2/{\rm d.o.f.}$ from $11.8$ to $4.4$. 
\end{itemize}
The $\chi^2/{\rm d.o.f.}$ remains large as the arrival time uncertainties are estimated without accounting for the impact of SWIMS.

\begin{figure*}
\includegraphics[angle=-90,width=\textwidth]{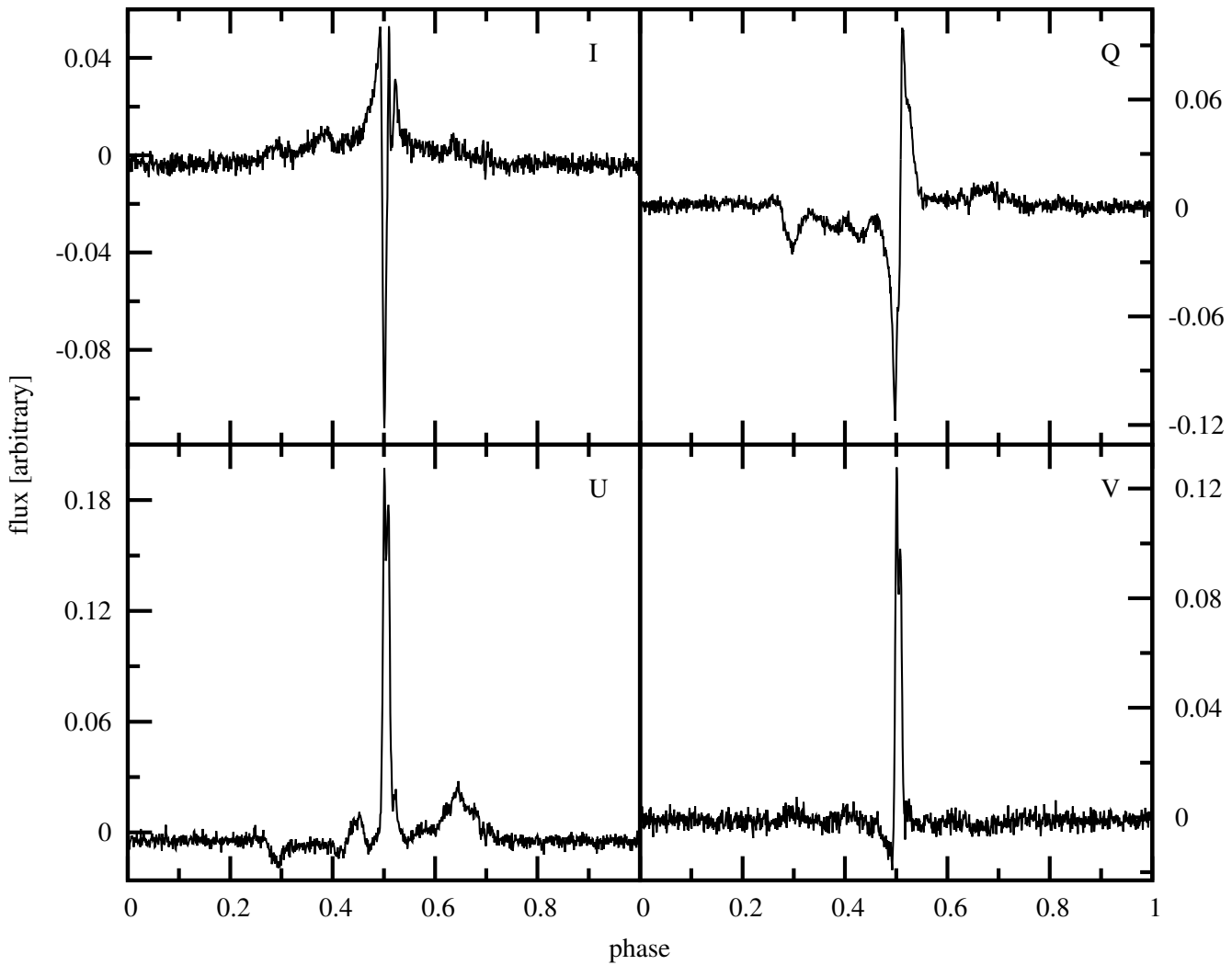}
\caption{First, most significant eigenvector, presented in the same way as the template profile in Fig. \ref{Fig::template_SWIMS_pol}, except we do not show the polarization angle for the eigenvector. 
Note that the flux in total intensity is smaller than flux in the polarized signal.
\label{Fig::evec_0}}
\end{figure*}

\begin{figure*}
\includegraphics[angle=-90,width=\textwidth]{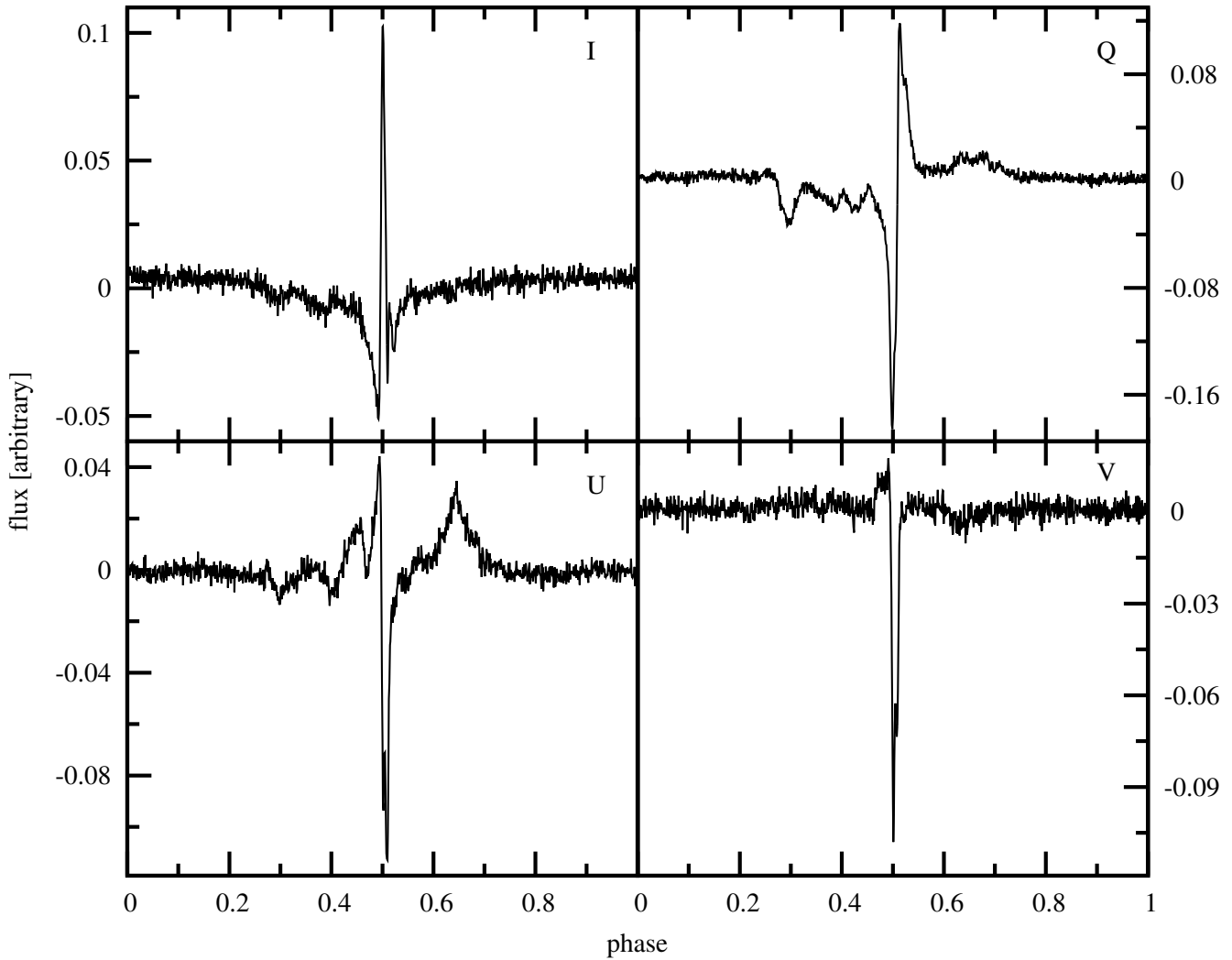}
\caption{The second eigenvector.
\label{Fig::evec_1}}
\end{figure*}

\begin{figure*}
\includegraphics[angle=-90,width=\textwidth]{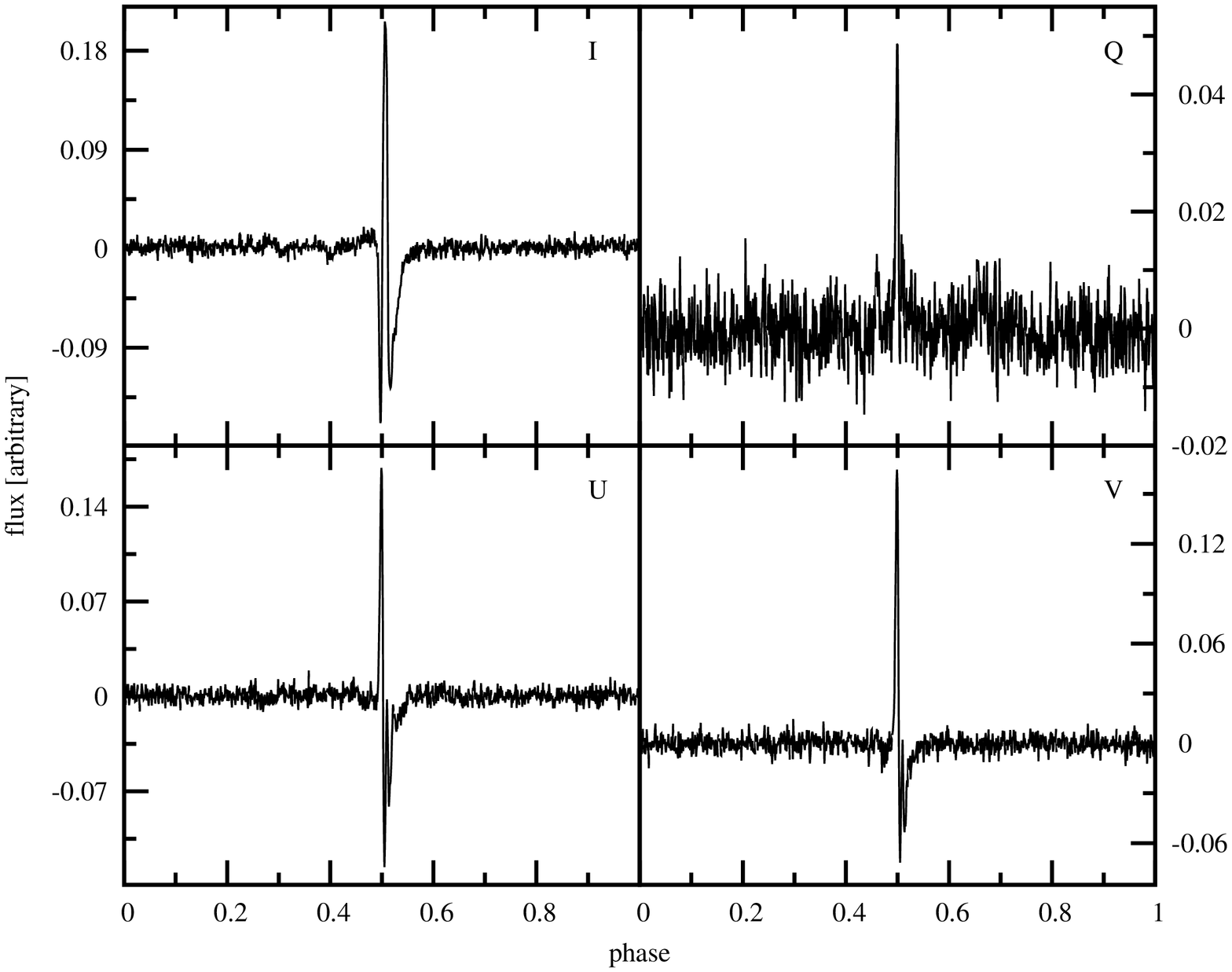}
\caption{The third eigenvector.
\label{Fig::evec_2}}
\end{figure*}

The first three, most significant eigenvectors are plotted in in Fig.~\ref{Fig::evec_0},~\ref{Fig::evec_1},~and~\ref{Fig::evec_2}. Each panel represents a different stokes parameter, from total intensity on top, via Q and U to V at the bottom. We note that the Stokes Q and U parameters in the first eigenvector are highly correlated and anti-correlated with the Stokes U and Q, respectively, of the template profile shown in Fig. \ref{Fig::template_SWIMS_pol}. The Stokes U parameter in the first eigenvector has an additional significant peak near the leading edge of the total intensity profile as compared to the Stokes U of the template profile. The majority of the detected pulse profile variance is located in the central parts of the profile.


\begin{figure}
\includegraphics[angle=0,width=\columnwidth]{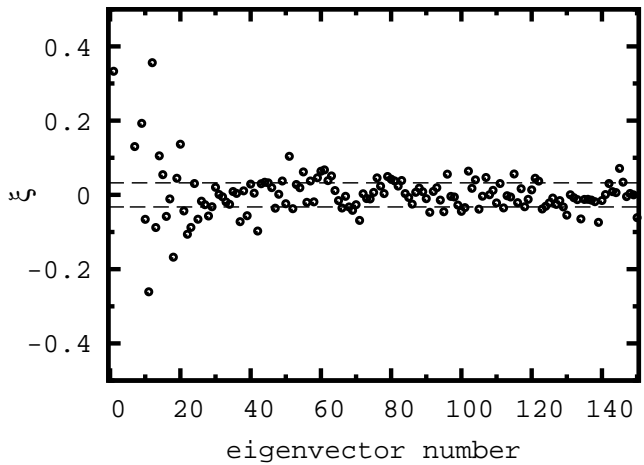}
\caption{Distribution of correlation coefficients between the arrival time residuals and projection on to eigenvectors for the actual observations, shown by open circles. The threshold level of three robust standard deviation estimate is marked with dashed lines. The correlation coefficients are highly scattered, thus requiring more stringent criteria for selection of the significant eigenvectors. \label{Fig::corel}}
\end{figure}

We plot the first 150 values of $\xi$ in Fig. \ref{Fig::corel}. In Paper I, we chose the number of significant eigenvectors by calculating a resistant and robust estimate of the standard deviation of $\xi$ and searched for three subsequent values of $\xi$ above three times the measured standard deviation. Applying the same criteria for automatic selection of the number of significant eigenvector results in 301 significant eigenvectors. This value prompted a closer investigation of the correlation coefficients $\xi$ and revealed that the automatic selection process does not work very well in this case. With a larger number of eigenvectors available, more stringent criteria are necessary or the number needs to be chosen by investigating the $\xi$ values manually. In this case, we arrive at a reasonable number of significant eigenvectors by increasing the required number of consecutive points above a chosen threshold to five.

\section{Discussion}
\label{Section::discussion}

We first discuss the details of the extended method and stress the differences between this new method and the one presented in paper~I. We then discuss the impact of the extended method on the timing precision of \psr\ and on the prospects of precision timing.

\subsection{Discussion of the method}
\label{Section::discussion_method}

The previously published method used pulse profiles resolved into $N_{\rm bin}$ phase bins but exploited only one stokes parameter. In the extension presented in this publication we employ all four stokes parameters with the same phase  resolution as before. This implies that in order for the covariance matrix to be fully determined we now need a minimum of $4N_{\rm bin}$ observations. Our conclusions from paper~I still hold and we briefly reiterate them here: a) when the number of available observations is lower than $4N_{\rm bin}$, the number of eigenvectors with non-null eigenvalues will be limited to the number of observations and care needs to be taken when choosing the number of significant eigenvectors; b) the number of required observations can be reduced by either gating the pulse profile or by reducing phase resolution; c) reducing the phase resolution of the pulse profile may wash away the temporal correlation of the noise which will render the presented methodology less efficient; d) in addition, reducing the phase resolution can lead to aliasing harmonics of spin frequencies of the observations as, by definition, we are dealing with high S/N observations; e) high number of observations is highly desirable in order to increase the S/N of the covariance matrix estimate and to ensure complete measurement of the SWIMS representations in the data; f) given a constant total integration time of all available observations, a larger number of separate sub-integrations will yield a higher S/N estimate of the covariance matrix.

As in paper~I, the eigenvectors are sorted by placing the corresponding eigenvalues in decreasing order such that the greatest variance in the data is associated with the first eigenvector and the eigenvectors corresponding to the fit-space are the last ones as all of the variance in the data in these dimensions is removed during the template fitting procedure. Despite the resulting monotonicity of the eigenvalues, the regression coefficients used in the ToA estimate correction scheme are not monotonic as the highest variance in profile does not directly translate into the highest bias of the ToA estimate. For this reason, applying multiple regression is crucial for a successful correction scheme and projections onto more than one eigenvector need to be used.

We note that care must be taken when making any physical interpretation of the eigenvectors, even though the detected pulse shape variations are likely to be pulsar intrinsic (see Paper I). This is due to the fact that our method calculates the eigenvectors of the covariance matrix assuming Euclidean geometry, i.e., the measured eigenvectors are orthogonal only in a classical sense. The stokes parameters span the Minkowski space and to allow any physical interpretations of the measured eigenvectors, their orthogonality must be ensured in the Minkowski space instead \citep[see, e.g.,][]{MinkowskiSVD}. Four dimensional Euclidian orthogonal transformations may not preserve the time-like interval of the stokes parameters. A prime example of this effect is visible in Fig. \ref{Fig::evec_0}, in which the polarized flux of the first eigenvector exceeds total intensity. As was the case in the method based on only the total intensity, the obtained eigenvectors are affected by the template matching process, this time with six degrees of freedom. This means that if one was to perform a simulation with a known pulse profile distortion inserted into data, the measured eigenvector would not correspond directly to the inserted noise (see Fig. 6 in paper~I and the relevant text).

The method presented here is superior to the technique presented in Paper I as the latter was limited primarily by any pulse shape variations correlated with the time derivative of the total intensity profile. In paper~I, the template matching procedure removes all signal projected on the fit-space. Variations that correlate with the time derivative of the template profile can introduce very significant bias into the ToA estimate without adding much power to the residual profile, as can be demonstrated by simulations. By including the information about the statistics of polarized noise, as presented in this work, some part of the information previously completely removed during the template fitting is retained in the Stokes Q, U and V parameters as these three parameters are not used to estimate ToA. Thus, shape variations in these parameters correlated with phase shift are not removed.

The covariance matrix constructed in the way described in \S \ref{Section::method} is related to the covariance matrix of the stokes parameters at a given phase bin as described by \citet{2009ApJ...694.1413V} in equation 28.  In \citet{2010ApJ...719..985V} the author generalised this matrix to include cross covariances between different polarized states. We note that the covariance matrix \textbfss{C} used here contains more information than either of the two above formulations as it contains information both about covariances between the stokes parameters at any phase bins as well as between different phase bins.

\subsection{Application to \psr}
\label{Section::discussion_application}

For comparison, the method presented in paper~I yields timing residual with an rms of $642$~ns, a 17 per cent reduction; and $\chi^2/{\rm d.o.f.}$ of 9.23. The improvement based on this method is somewhat worse than for the dataset used in paper~I, most likely due to the lower S/N or residual 2-bit distortions of the observations used in this publication. By exploiting the information available in the polarization of the incident radiation we are able to improve the results of the bias removal from ToA estimates by more than a factor of two compared with the previously published approach. We note that despite the significant improvement in the attained rms timing residual, the results may vary greatly for different pulsars depending on their pulse profile and the nature of SWIMS.

As we are employing the information contained in the polarized incident radiation, it is useful to compare the rms timing residual obtained using the methodology presented in this paper with other methods that use not only the total intensity but the other three stokes parameters as well. Two such methods are the matrix template matching \citep{2006ApJ...642.1004V} and timing of the invariant interval pulse profile \citep{2000ApJ...532.1240B}. The first allows us to achieve higher timing precision by exploiting the fact that for some pulsars the polarized signal contains more power in the higher harmonics of the Fourier transform of the pulse profile and can compensate for some of the systematic errors introduced during the calibration. The latter minimises the errors arising from polarimetry related errors by calculating the invariant interval and is for example typically used for the precision timing of \psr\ within the Parkes Pulsar Timing Array project \citep{Manchester2012}.

The ToAs derived using these two methods have an rms value of 588 and 771 nanoseconds and $\chi^2/{\rm d.o.f.}$ of 4.9 and 5.2, respectively. The $\chi^2$ value is significantly better than the corresponding value for ToAs derived from total intensity as calculation of the invariant interval significantly reduces the S/N of the data for highly polarized pulsars, such as \psr, thus making the bias introduced by SWIMS less obvious by increasing the uncertainties in the estimated ToAs. Furthermore, systematic errors might be present in the invariant interval based on observations with CPSR2 due to the automatic gain control used by this baseband system.

It is possible to correlate the ToA residuals derived from matrix template matching with the residual profiles expressed in the basis spanned by the eigenvectors to obtain an rms timing residual of $449$~ns and $\chi^2/{\rm d.o.f.}$ of $2.9$. The rms timing residual is similar to the value presented in \S \ref{Section::results} and 23.5 per cent lower than the input value. The $\chi^2/{\rm d.o.f.}$ value is smaller than presented in \S \ref{Section::results} as the ToA uncertainty estimate based on matrix template matching is more realistic because it is the formal least squares uncertainty, which takes into account the covariances between the fit parameters.

\begin{figure}
\includegraphics[angle=0,width=\columnwidth]{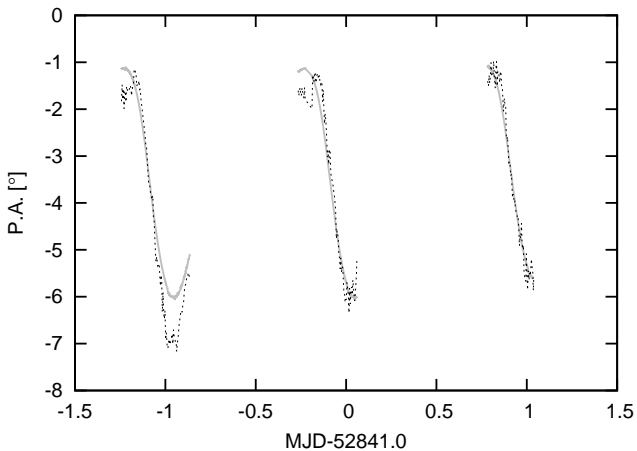}
\caption{The position angle over three chosen days of observations. Grey solid line shows the combined predictions of IRI and IGRF while black dashed line corresponds to the measurements of the absolute position angle. We note that while the overall agreement is very good, at some epochs there are significant differences between the models and data. \label{Fig::ionosphere}}
\end{figure}

The method presented here is very sensitive to polarimetric calibration of the data and any imperfections in calibration will be detectable by presence of characteristic signals in the eigenvectors. Any effect that affects the polarization of the pulsar should be detectable using this method. The fact that the first eigenvector for Stokes Q and U are correlated with the Stokes U and Q parameters in the template, respectively, can be explained by Faraday rotation the Earth's ionosphere. Based on the International Reference Ionosphere\footnote{http://iri.gsfc.nasa.gov} (IRI) and International Geomagnetic Reference Field\footnote{http://www.ngdc.noaa.gov/IAGA/vmod/igrf.html} (IGRF) models, we derived a strong correlation with correlation coefficient of $0.83$ between the relative position angle derived from 5-minute integrations using matrix template matching \citep{2006ApJ...642.1004V} and the predicted position angle change induced by the ionosphere. These calculations were performed using software developed by \citet{2011Ap&SS.335..485Y} and are shown in Fig.~\ref{Fig::ionosphere}. Note that the ionosphere was quite stable during the week of observations, however the calibration solutions were derived at different times of the day, corresponding to different angles between the line of sight and the Earth's magnetic field. The coincidence of a spike in Stokes I, U and V and the mixing of Stokes Q and U is most likely due to the projection of the $4N_{\rm bin}$-dimensional space on to the subspace that has the six-dimension fit space removed. It does not imply that these two types of variation are correlated.

For pulsar timing experiments, it is often desirable to derive ToAs based on hour-long observations. However, the temporal correlation of the mixing of Stokes Q and U implies that the covariance matrix may no longer scale as the inverse of integration time. This in turn renders the predictor derived from one time-scale inapplicable to other time-scales. In principle, without red noise present and with ionosphere related effects removed, we should achieve rms timing residual of 30~ns with one-hour integrations with the dataset presented. Note that this is comparable to what modern observing systems with four times larger bandwidth achieve \citep[][]{Manchester2012}. To render the predictors useful at all scales, calibration solutions need to be derived at higher time resolution, or ultimately the state of the ionosphere can be monitored in near real-time using the dual-frequency Global Positioning System satellites \citep[see e.g.,][]{2008AdSpR..42..720G}. Both these options should allow to remove the mixing of Stokes Q and U and enable higher precision experiments. 

\psr\ is about 100 times brighter than other known MSPs and, as noted in Paper I, unless this object is unusual, we can expect that SWIMS will limit the timing precision of other MSPs with the next generation of radio telescopes such as the Square Kilometre Array or Five hundred metre Aperture Spherical Telescope. Based on the expected collecting areas and system temperatures of these instruments \citep{2006ScChG..49..129N,2010SPIE.7733E..35S}, one can predict the timing precision of future timing experiments. By doing so, \citet{2011MNRAS.417.2916L} concluded that the timing precision of MSPs will be greatly improved with these instruments; however, it will likely be limited by the impact of SWIMS on the ToA estimation. This warrants the continued effort into development of appropriate algorithm to deal with SWIMS, such as the one presented in this work.

\section{Conclusion}
\label{Section::conclusions}

A solution to the long-standing problem of failing to achieve the expected timing precision for \psr\ was presented in paper~I. In the same work, a method for improving the precision of the timing of this pulsar was presented. Here, we address one limitation of the previous method and subsequently attain a reduction of more than 40 per cent of the rms timing residual, more than a factor of two better than the method published in paper~I. This corresponds to reduction of observing time necessary to achieve a given precision by almost a factor of three. Thus this method is likely to be very important for many pulsars when observed with the next generation of radio telescopes. The rms timing residual achieved with the method presented here is lower than can be attained with any other available method, including matrix template matching and timing of invariant interval pulse profiles.  Currently, the applicability of this method, and thus the timing precision of \psr, is limited by ignoring the Earth's ionosphere in the data analysis, thus providing a natural next step in the pursuit of ultimate timing precision that remains to be addressed in future work. After correcting the influence of Earth's ionosphere it will be possible to achieve an rms timing residual of about $30$~ns in one hour integration. Improving the timing precision for a number of pulsars will impact a number of astrophysical experiments, in particular the detection of gravitational waves with pulsar timing arrays.

\section*{Acknowledgements}

The Parkes Observatory is part of the Australia Telescope National Facility which is funded by the Commonwealth of Australia for operation as a National Facility managed by CSIRO. We thank the staff at Parkes Observatory for technical assistance during observations. The authors express their gratitude to the anonymous referee and Joris Verbiest for very useful comments on the manuscript, Richard Manchester for making the code calculating the position angle based on IRI and IGRF available, and Jim Cordes and Ryan Shannon for early discussion on the topic of principal component analysis in the context of precision pulsar timing. This work was supported by Australian Research Council grant \# DP0985272. 

\newcommand{\apj}{ApJ}
\newcommand{\aj}{AJ}
\newcommand{\apjs}{ApJS}
\newcommand{\apjl}{ApJ Lett}
\newcommand{\nat}{Nature}
\newcommand{\aap}{A\&A}
\newcommand{\prc}{Phys.~Rev.~C}
\newcommand{\prd}{Phys.~Rev.~D}
\newcommand{\physrev}{Phys. Rev.}
\newcommand{\mnras}{MNRAS}
\newcommand{\nar}{New Astronomy Reviews}
\newcommand{\pasp}{PASP}
\newcommand{\pasj}{PASJ}
\newcommand{\apss}{ApSS}
\newcommand{\aapr}{AAPR}
\newcommand{\aaps}{A\&AS}
\newcommand{\physrep}{Phys. Rep.}
\newcommand{\sovast}{Soviet Astron.}
\newcommand{\pasa}{Publ. Astron. Soc. Aust.}

\footnotesize{
\bibliographystyle{mn2e}
\bibliography{SWIMS_stokes}
}

\label{lastpage}

\end{document}